\documentstyle[12pt]{article}

\def\be{\begin{equation}}
\def\ee{\end{equation}}
\def\bdi{\begin{displaymath}}
\def\edi{\end{displaymath}}
\def\br{\begin{eqnarray}}
\def\er{\end{eqnarray}}
\def\no{\nonumber}
\def\o{\over}

\def\u2{\mid u\mid^2}

\def\RR{{\rm I\kern-.1567em R}}                              
 \def\CC{{\rm C\kern-4.7pt                                    
 \vrule height 7.7pt width 0.4pt depth -0.5pt \phantom {.}}} 
 \def\ZZ{{\sf Z\kern-4.5pt Z}}                                

\begin{document}

\begin{titlepage}
\vspace*{-2 cm}
\noindent

\vskip 1cm
\begin{center}
{\Large\bf Integrable subsystem of Yang--Mills dilaton theory  }
\vglue 1  true cm
C. Adam$^{1a}$,   J. S\'anchez-Guill\'en$^{1b}$,  and 
A. Wereszczy\'nski$^{2c}$
\vspace{1 cm}

$^1${\footnotesize Departamento de F\'\i sica de Part\'\i culas,\\
Facultad de F\'\i sica,
Universidad de Santiago, \, \, and \\
Instituto Galego de Fisica de Altas Enerxias (IGFAE) \\
E-15782 Santiago de Compostela, Spain} \\ ${}$ \\
$^2${\footnotesize Institute of Physics, Jagiellonian University, \\
Reymonta 4,  30-059 Krak\'ow, Poland}

\vspace{1 cm}

\medskip
\end{center}

\normalsize
\vskip 0.2cm

\begin{abstract}
With the help of the Cho--Faddeev--Niemi--Shabanov decomposition of the
SU(2) Yang--Mills field, we find an integrable subsystem of SU(2)
Yang--Mills theory coupled to the dilaton. Here integrability means
the existence of infinitely many symmetries and infinitely many conserved
currents. Further, we construct infinitely many static solutions of this
integrable subsystem. These solutions can be identified with certain limiting
solutions of the full system, which have been found previously in the context
of numerical investigations of the Yang--Mills dilaton theory. In addition,
we derive a Bogomolny bound for the integrable subsystem and show that our
static solutions are, in fact, Bogomolny solutions. This explains
the linear growth of their energies with the topological charge, which has
been observed previously. Finally, we discuss some generalisations.

\end{abstract}

\vfill

$^a${\footnotesize adam@fpaxp1.usc.es} 

$^b${\footnotesize joaquin@fpaxp1.usc.es}

$^c${\footnotesize wereszczynski@th.if.uj.edu.pl}

\end{titlepage}
\section{Introduction }
Pure Yang--Mills theory does not allow for static finite energy solutions, 
as follows from the scale invariance of this theory. One simple way to
circumvent this obstacle is to couple the Yang--Mills Lagrangian to the
dilaton field \cite{MaLa1}, \cite{Biz}. The corresponding Lagrangian reads
\be \label{YM-dil}
{\cal L} = \frac{1}{4} \left( 2 \partial^\mu \xi \partial_\mu \xi -
e^{-2\kappa \xi}  F^{a\mu\nu} F^a_{\mu\nu} \right) 
\ee
where $F^a_{\mu\nu}$ is the SU(2) Yang--Mills field strength, and we choose 
units such that the gauge field coupling is equal to one.
Further, $\xi$ is the dilaton field, and $\kappa$ is the dilaton coupling
constant. A direct physical application of the above theory would require the
inclusion of further interactions (e.g., the coupling to gravity for a
low-energy effective theory of string theory), but in this paper we 
deal with the theory given by (\ref{YM-dil}) for the sake of simplicity.

Static solutions of the Yang--Mills dilaton theory were first discussed in
\cite{MaLa1}, \cite{Biz}, and in those papers infinitely many unstable, 
sphaleron type
solutions were found numerically within a spherically symmetric
ansatz. Further, in \cite{Biz} 
an effectively abelian solution within the spherically
symmetric ansatz was found analytically, which provided a
limiting case for the numerical solutions.
A similar analysis was performed in \cite{KuKl1}, this time for an ansatz with
only cylindrical symmetry. Again, sequences of infinitely many sphaleron 
solutions, labelled by a winding number $m$, were
found numerically, and infinitely many effectively abelian limiting solutions
characterised by the same winding number were constructed
analytically. Further, it was observed that the energies of the limiting
solutions grow linearly with the winding number $m$. 

We shall find that all
these effectively 
abelian limiting solutions belong to an integrable submodel of
Yang--Mills dilaton theory characterised by infinitely many symmetries and
infinitely many conserved currents, 
and that in this submodel there exists a Bogomolny
bound, explaining the linear growth of energy with winding number.
Recently, further solutions with only discrete symmetries (rotational
symmetries of platonic bodies) have been investigated numerically in
\cite{KuKl2}.

Our paper is organised as follows. In Section 2, we briefly review the
Cho--Faddeev--Niemi--Shabanov (CFNS) decomposition of the 
SU(2) Yang--Mills field. Then we use this decomposition to define a submodel
of Yang--Mills dilaton theory
by restricting the decomposition fields, and show that this submodel is
integrable in the sense that it has infinitely many symmetries and infinitely
many conserved currents. In Section 3 we use a separation of variables ansatz
for static configurations of the integrable submodel and show that the
ordinary differential equations (ODEs) obtained in this way can be solved by
quadratures. The resulting solutions are precisely the limiting solutions of
\cite{Biz}, \cite{KuKl1} with spherical and cylindrical symmetry, 
respectively. In Section 4 we show that there exists a Bogomolny
bound for the integrable submodel, and demonstrate
that the solutions of Section 3 are, in
fact, Bogomolny solutions. In Section 5 we discuss some generalisations
allowing for additional solutions of the integrable submodel, and 
introduce some more
general integrable Lagrangians, which are no longer related to the Yang--Mills
dilaton theory, but still have static field equations which can be solved by
quadratures. Section 6 contains our conclusions.  In the appendix we prove
that the two possible ways to derive the field equations of the integrable
subsystem (restriction to the subsystem already in the
Lagrangian, on the one hand,
or insertion of the restriction into the field
equations of the full Yang--Mills dilaton theory,
on the other hand) really lead to the same
field equations. 

\section{The integrable subsystem}

The Cho--Faddeev--Niemi--Shabanov (=CFNS) 
decomposition (see, e.g., \cite{Cho1} - \cite{Gies1}) 
expresses the gauge field as a sum of three terms
like follows,
\be \label{CFNS-dec}
A^a_\mu = n^a C_\mu + \epsilon^{abc} n^b_\mu n^c + W^a_\mu ,
\ee
where $ n^a = (n^1 ,n^2 ,n^3)$ is a unit vector in $SU(2)$ color space,
$C_\mu$ is an abelian gauge potential in the $ n^a$ direction in color
space, and $ W^a_\mu$ is orthogonal to $ n^a$ in color space,
$n^a W^a_\mu =0$. 
Further, $n^a_\mu \equiv \partial_\mu n^a$ (spacetime indices on scalar fields
will always denote partial derivatives).  In the sequel,
we shall refer to the three terms at the r.h.s. of Eq. (\ref{CFNS-dec}) as the
$C$-term, the $n$-term, and the $W$-term, respectively.
 
In order to guarantee the correct gauge transformation properties
\br
\delta n^a &=& \epsilon^{abc} n^b \alpha^c \no \\
\delta W^a_\mu  &=& \epsilon^{abc} W^b_\mu  \alpha^c \no \\
\delta C_\mu &=& n^a \alpha^a_\mu 
\er 
under the gauge transformation
\be
\delta A^a_\mu = (D_\mu \alpha )^a \equiv \alpha^a_\mu + \epsilon^{abc}
A^b_\mu \alpha^c ,
\ee
the following constraint has to be imposed,
\be \label{CFNS-const}
\partial^\mu W^a_\mu + C_\mu \epsilon^{abc} n^b W^c_\mu + n^a W^b_\mu n^b_\mu
\equiv 0.
\ee 
In addition, this constraint makes that the number of degrees of freedom of
the gauge field and of the decomposition match.

We now want to restrict to a specific class of gauge fields where we set 
the $W$-term (the valence field) equal to zero
\be
W^a_\mu =0,
\ee
i.e., we assume the restriction
\be \label{CFNS-res}
\hat A^a_\mu = n^a C_\mu + \epsilon^{abc} n^b_\mu n^c .
\ee
This restriction is gauge invariant, because $W^a_\mu$ transforms
homogeneously under gauge transformations. Further, the restricted 
potential $\hat A^a_\mu$ still transforms like a SU(2) gauge potential
under gauge transformations. The corresponding field strength is,
nevertheless, abelian, therefore the above restriction also provides a gauge
invariant definition of the abelian projection \cite{Cho3}.
Note that this abelian projection is also compatible with the constraint 
(\ref{CFNS-const}).

In a first step, we further restrict to gauge potentials which are
solely described by the unit vector $n^a$, i.e., we set
\be \label{CW-restr}
C_\mu =0, \quad W^a_\mu =0.
\ee
This choice is no longer gauge invariant and also further reduces the number
of degrees of freedom. Later on we shall allow for the more general gauge
potentials (\ref{CFNS-res}) with $C_\mu \not= 0$, and we will also discuss in
more detail the issue of gauge transformations, see Section 5.    
It is sometimes 
assumed that $n^a$ describes the low-energy degrees of freedom of the
Yang--Mills field (which was in fact one motivation for the decomposition),
but we shall not be concerned with this here.  

Inserting the decomposition (\ref{CFNS-dec}) with the restriction 
(\ref{CW-restr}) into the Yang--Mills dilaton Lagrangian (\ref{YM-dil}) we
arrive at the Lagrangian
\be \label{int-lag-0}
{\cal L} = \frac{1}{4} \left( 2 \partial^\mu \xi \partial_\mu \xi -
e^{-2\kappa \xi}  H^{a\mu\nu} H^a_{\mu\nu} \right) 
\ee
where
\be
H^a_{\mu\nu} \equiv \epsilon^{abc}n^b_\mu n^c_\nu = n^a H_{\mu\nu},
\quad H_{\mu\nu} \equiv \epsilon^{abc}n^a n^b_\mu n^c_\nu.
\ee

Remark: we shall use this Lagrangian (\ref{int-lag-0}) in Section 3 to
derive the corresponding Euler--Lagrange equations. Here, of course,
the question arises whether these equations are really equivalent to the
original Euler--Lagrange equations of the Yang--Mills dilaton theory after the
decomposition is inserted into these original equations. We prove in the
appendix that this is indeed the case. 

For later convenience we prefer to replace the three-component unit vector
field $ n^a$ by the complex scalar field $u$ via stereographic projection,
 \be
 n^a = {1\o {1+\mid u\mid^2}} \, ( u+\bar u , -i ( u-\bar u ) ,  
1-u\bar u ) \; ;
\qquad
u  = \frac{n_1 + i n_2}{1 + n_3}.
\label{stereo}
\ee
Then we get for the above Lagrangian (\ref{int-lag-0})
\be \label{H_munu}
H_{\mu\nu} = 2i \frac{u_\mu \bar u_\nu - u_\nu \bar u_\mu}{(1+u\bar u)^2}
\ee
and 
\be \label{int-lag}
{\cal L} = \frac{1}{2}   \partial^\mu \xi \partial_\mu \xi - 2
e^{-2\kappa \xi} \frac{(u^\mu \bar u_\mu )^2 - (u_\mu)^2 (\bar u_\nu)^2}{
(1+u\bar u)^4} .
\ee
This Lagrangian is integrable in the sense that it has infinitely many
symmetries and infinitely many conserved currents. Indeed, it is of the type
c) of Table 3 of Ref. \cite{vol-pres} and, therefore, has the infinitely
many conserved currents 
\be \label{J-tilde-G}
J^{\tilde G}_\mu = i \tilde G_a (u \pi_\mu - \bar u \bar \pi_\mu )
\ee
where 
\be 
\pi_\mu \equiv \frac{\partial \cal L}{\partial u^\mu} \, ,\quad
\bar \pi_\mu \equiv \frac{\partial \cal L}{\partial \bar u^\mu}
\ee
are the usual canonical four-momenta and $\tilde G=\tilde G(a) $ 
($a\equiv u\bar u$)
is an arbitrary real function of its argument. 

Remark: the Lagrangian (\ref{int-lag}) has, in fact, an even larger symmetry.
It belongs to a rather special class which has not been classified explicitly
in Ref. \cite{vol-pres}. Indeed, this Lagrangian belongs both to type c) and
to type b) of Table 3 of Ref. \cite{vol-pres}, i.e., it depends on the
target space variable $a\equiv u\bar u$  solely via the target space
metric function $g$,
\be
{\cal L} ={\cal F}(g^2 c,d)
\ee
where
\be
g \equiv   e^{-\kappa \xi} (1+a)^{-2} \, \qquad a\equiv u\bar u
\ee
\be
c\equiv (u^\mu \bar u_\mu )^2 - (u_\mu)^2 (\bar u_\nu)^2 \, ,\quad
d\equiv \xi^\mu \xi_\mu .
\ee
Further, the metric function is of the product form 
\be
g = g^{(1)} (a)g^{(2)}(\xi) .
\ee
As a consequence, the Lagrangian (\ref{int-lag}) has the infinitely many
conserved currents
\be \label{J-cal-G}
J^{ \cal G}_\mu = \frac{i}{g^{(1)}(a)} [{\cal G}_{\bar u} \pi_\mu - {\cal G}_u 
\bar \pi_\mu ]
\ee
where ${\cal G}={\cal G}(u,\bar u)$ 
is an arbitrary real function of its arguments.
From a geometric point of view the existence of these conserved currents is
quite obvious, because they are just the Noether currents of the
area-preserving diffeomorphisms on the target two-sphere which is spanned
by the complex field $u$, and these are symmetries of the above Lagrangian,
see, e.g., \cite{FR1}, \cite{BF1}, \cite{ASG4}, \cite{ASGW1}. 

\section{Static solutions}

The energy functional for static configurations which corresponds to the
Lagrangian (\ref{int-lag}) reads
\be \label{en-int-mod}
E = \int d^3 {\bf r} \left( \frac{1}{2}   \nabla \xi \cdot \nabla \xi + 2
e^{-2\kappa \xi} \frac{(\nabla u \cdot \nabla \bar u )^2 - 
(\nabla u)^2 (\nabla \bar u)^2}{(1+u\bar u)^4} \right) .
\ee
Momentarily we want to switch to the more general class of energy functionals
\be \label{en-int-mod-gen}
E = \int d^3 {\bf r} \left( \frac{1}{2}   \nabla \xi \cdot \nabla \xi +
2  G^{(1)}(a) G^{(2)}(\xi )
[(\nabla u \cdot \nabla \bar u )^2 - 
(\nabla u)^2 (\nabla \bar u)^2 ] \right) 
\ee
because this general class has exactly the same integrability properties and
may be solved in exactly the same way. Later we will specialise to the case
\br
g^{(1)}(a) &=&  (1+ a)^{-2} \, , 
\quad G^{(1)} \equiv \left( g^{(1)} \right)^2
\no \\
g^{(2)}(\xi) &=& e^{-\kappa \xi} \, , \quad G^{(2)} \equiv 
\left( g^{(2)} \right)^2
\er
when it is needed.

First of all, let us observe that the energy functional 
(\ref{en-int-mod-gen}) is the sum of
a term which is quadratic in first derivatives and another term which is
quartic. Therefore, the Derrick criterion does not rule out the existence of
static finite energy solutions, and we will indeed find that they exist.  
The Euler--Lagrange equation for the variation w.r.t. $\bar u$ is
\be \label{EL-u}
\nabla \cdot \left( g^{(1)}(a) G^{(2)}(\xi) \vec K\right) =0
\ee
where
\be
\vec K \equiv (\nabla u \cdot \nabla \bar u)  \nabla u - (\nabla u)^2 
\nabla \bar u
\ee
(observe the appearence of $g^{(1)} \equiv \sqrt{G^{(1)}}$ in the equation).
The Euler--Lagrange equation for the variation w.r.t. $\xi$ is
\be \label{EL-xi}
\Delta \xi = 2 G^{(1)} G^{(2)}_\xi [
(\nabla u \cdot \nabla \bar u )^2 - 
(\nabla u)^2 (\nabla \bar u)^2 ] .
\ee

Next, we introduce spherical polar coordinates $\vec r =(r\sin \theta 
\cos \varphi ,r\sin \theta \sin \varphi ,r\cos \theta ) $ and the
corresponding frame of unit basis vectors $(\hat e_r ,\hat e_\theta ,
\hat e_\varphi )$, and we employ the ansatz
\be \label{ansatz}
\xi = \xi (r) \, ,\quad u = v(\theta ) e^{im\varphi} .
 \ee
Then the Euler--Lagrange equation (\ref{EL-u}) simplifies to
\be \label{EL-u-1}
\nabla \cdot \left( g^{(1)}(a) \vec K\right) =0
\ee
because $\nabla \xi \cdot \vec K=0$. With 
\be
\nabla u = \frac{1}{r} e^{im\varphi } \left( v_\theta \hat e_\theta
+ \frac{imv}{\sin \theta } \hat e_\varphi \right) 
\ee
and
\be
\vec K = \frac{2}{r^3}e^{im\varphi} \left( \frac{m^2 v^2 v_\theta }{\sin^2
\theta }\hat e_\theta + \frac{imvv_\theta^2 }{\sin \theta } \hat e_\varphi
\right) 
\ee
and using that $g^{(1)}(a(\theta))$ is a function of $\theta$ only, we find
\be
\nabla \cdot \left( g^{(1)}(a) \vec K\right) =
\frac{2m^2 e^{im\varphi}}{r^4 \sin \theta } \partial_\theta
\left( \frac{g^{(1)} vv_\theta}{\sin \theta }  \right) \equiv 0
\ee
and, therefore, the first trivial integral 
\be \label{first-int-theta}
\frac{g^{(1)} vv_\theta}{\sin \theta } =\mu ={\rm const.} .
\ee
The further evaluation depends on the explicit form of $g^{(1)}$.
Choosing $g^{(1)}=  (1+a)^{-2} = (1+v^2)^{-2}$ we get
\be
\frac{V_\theta}{(1+V)^2} = \mu \sin \theta \, \qquad V\equiv v^2
\ee
and the solution
\be
V=\frac{1-\mu \cos \theta +\lambda}{\mu \cos\theta -\lambda}
\ee
where $\lambda$ and $\mu$ are two integration constants. 
The integration constants
are fixed by the requirement that $u$ should be a genuine map $S^2 \to S^2$.
This requires that the modulus $v$ of $u$ should cover the whole positive 
real semiaxis, i.e., 
\be
V(0)=0 \, ,\quad V(\pi )=\infty .
\ee
This leads to
\be \label{bound-theta}
\lambda =-\frac{1}{2} \, ,\quad \mu = \frac{1}{4}
\ee
and to
\be \label{sol-theta}
V=\frac{1-\cos \theta}{1+\cos\theta} \, ,\qquad
u=\tan\frac{\theta}{2}e^{im\varphi}  .
\ee
Indeed, the corresponding 
$u$ describes a map $S^2 \to S^2$ with winding number $m$.   The resulting
field strength $H_{jk}$ just describes the abelian magnetic monopole 
with charge $m$. Indeed, for the hodge dual vector
\be
H^i = \epsilon^{ijk}H_{kj}
\ee
we get
\be
\vec H = \frac{m}{r^2}\hat e_r .
\ee
The magnetic charge $m \in \ZZ$ is quantised by the topological nature of $u$.

In order to solve the Euler-Lagrange equation (\ref{EL-xi}), we first need 
the expression
\be
(\nabla u \cdot \nabla \bar u )^2 - 
(\nabla u)^2 (\nabla \bar u)^2 = \frac{4m^2 v^2 v_\theta^2}{r^4 \sin^2 \theta}
\ee
for the ansatz (\ref{ansatz}). We find for the Euler-Lagrange 
equation (\ref{EL-xi})
\be
\frac{1}{r^2} \partial_r (r^2 \partial_r \xi )
=\frac{8m^2 v^2 v_\theta^2}{r^4 \sin^2 \theta} \left( g^{(1)} \right)^2
G^{(2)}_\xi  = \frac{8m^2 \mu^2}{r^4}  G^{(2)}_\xi
\ee
where we used the first integral (\ref{first-int-theta}). With the new
variable $s=r^{-1}$ we get
\be
\xi_{ss} = 8m^2 \mu^2 G^{(2)}_\xi
\ee
and, upon multiplication with $\xi_s$, the first integral
\be \label{first-int-r}
\xi_s^2 = 16 m^2 \mu^2 G^{(2)} + \tilde \lambda
\ee
where $\tilde \lambda$ is an integration constant. 
This expression may be easily
integrated, after taking the square root, and results in
\be
s+s_0 = \int \frac{d\xi }{\sqrt{16 m^2 \mu^2 G^{(2)} +\tilde \lambda }} ,
\ee
and $s_0$ is another integration constant. For a further evaluation one
has to choose an explicit function for $G^{(2)}$. Choosing $G^{(2)}
=e^{-2\kappa \xi}$, and $\mu =(1/4)$, we get
\be
s+s_0 = \int \frac{d\xi }{\sqrt{m^2  e^{-2\kappa \xi} +\tilde \lambda }} .
\ee 
For finite energy solutions one has to choose $\tilde \lambda =0$, 
in which case
the integral is trivial and has the solution
\be 
s+s_0 = \frac{e^{\kappa \xi}}{\kappa |m|} \quad \Rightarrow \quad
\xi = \frac{1}{\kappa} \ln [|m|\kappa (s+s_0)] .
\ee
For a further evaluation we have to impose boundary conditions. We require that
$\xi$ covers the whole positive real semi-axis which implies that
\be \label{bound-xi}
\xi (s=0) =0 \, , \quad \xi (s=\infty) =\infty \qquad \Rightarrow \qquad
s_0 =\frac{1}{\kappa |m|}
\ee
and, therefore
\be \label{sol-r}
\xi = \frac{1}{\kappa} \ln [|m|\kappa s+ 1] .
\ee
These solutions are precisely the limiting solutions (in the limit where the
number of nodes of a certain ansatz function in the numerical analysis
goes to infinity)
which have been found in
\cite{Biz} (for $m=1$) and in \cite{KuKl1} (for general $m$). 

For the energy we find (remember $s\equiv r^{-1}$)
\br
E &=& \int d^3 {\bf r} \left( \frac{1}{2}   \nabla \xi \cdot \nabla \xi + 2
e^{-2\kappa \xi} \frac{(\nabla u \cdot \nabla \bar u )^2 - 
(\nabla u)^2 (\nabla \bar u)^2}{(1+u\bar u)^4} \right) \no \\
&=& \frac{1}{2}\int d^3 {\bf r} \left( \xi_r^2 + \frac{m^2}{r^4} 
e^{-2\kappa \xi} \right) \no \\
&=& \frac{1}{2} 4\pi \int dr r^2 \left( \frac{m^2}{(|m| \kappa r +r^2)^2}
+ \frac{m^2}{r^4} \frac{1}{(|m| \kappa r^{-1} +1)^2} \right) \no \\
&=& 4\pi m^2 \int \frac{dr}{(|m| \kappa +r)^2} = \frac{4\pi |m|}{\kappa} . 
\label{energy}
\er 
Therefore, the energy is linear in the topological charge, as was already
observed in \cite{KuKl1}. This gives rise to 
the question whether there exists a
Bogomolny type bound in the integrable submodel. And that is indeed the case,
as we shall see in the next section.

\section{The Bogomolny bound}
We introduce the vector
\be
\vec H = 2i\frac{(\nabla u) \times (\nabla \bar u)}{(1+u\bar u)^2}
= \frac{1}{2} \epsilon^{abc} n^a (\nabla n^b )\times (\nabla n^c)
\ee
and express the energy functional like
\br
E &=& \frac{1}{2} \int d^3 {\bf r} [(\nabla \xi )^2 + G^{(2)}(\xi) \vec H^2 ]
\no \\
&=& \frac{1}{2} \int d^3 {\bf r} (\nabla \xi - g^{(2)} \vec H )^2 
+ \int d^3 {\bf r} g^{(2)} \nabla \xi \cdot \vec H \no \\
&\ge & \int d^3 {\bf r} g^{(2)} \nabla \xi \cdot \vec H \, \equiv
\, E_{\rm Bog}.
\label{bogol-ineq}
\er
Therefore, the Bogomolny equation is
\be \label{bogol-eq}
\nabla \xi - g^{(2)} \vec H =0.
\ee
We now show that our solutions (\ref{sol-theta}), (\ref{sol-r}) for the
ansatz (\ref{ansatz}) obey this Bogomolny equation. For the ansatz we have 
$\nabla \xi = \xi_r \hat e_r$ and 
\be
\vec H = \frac{4m}{r^2 \sin \theta}\frac{vv_\theta }{(1+v^2)^2} \hat e_r
= \frac{4m}{r^2 }\frac{g^{(1)}vv_\theta }{\sin\theta} \hat e_r
=\frac{m}{r^2} \hat e_r
\ee
where we used (\ref{first-int-theta}) and $\mu^{-1} =4$, see 
(\ref{bound-theta}). The Bogomolny equation now becomes
\be
\xi_r = \frac{m}{r^2}g^{(2)}
\ee
or, after the variable change $s=r^{-1}$ and squaring of the resulting
expression,
\be
\xi_s^2 = m^2 G^{(2)}
\ee
which is exactly the first integral (\ref{first-int-r}) for the special
choice $\tilde \lambda =0$ of the integration constant $\tilde \lambda$.
This is, of course, consistent with our remark that finite energy solutions
require  $\tilde \lambda =0$. In short, the static solutions of the last
section indeed obey the Bogomolny equation (\ref{bogol-eq}).

It remains to evaluate the Bogomolny energy in the Bogomolny inequality
(\ref{bogol-ineq}). We get
\be
E_{\rm Bog} = 
\int d^3 {\bf r} g^{(2)} \nabla \xi \cdot \vec H = 
\int d^3 {\bf r}  \nabla \tilde g \cdot \vec H
\ee
where $\tilde g$ obeys
\be
\tilde g_\xi =g^{(2)} (\xi)
\ee
Specifically, for our model $ g^{(2)} =e^{-\kappa \xi}$ and, therefore,
\be
E_{\rm Bog} =
-\frac{1}{\kappa }\int d^3 {\bf r}  (\nabla e^{-\kappa \xi})  \cdot \vec H .
\ee
This we want to compare now with an expression for the winding number 
$Q$ of a
map $\RR^3_0 \to S^3$, where $\RR^3_0$ is one-point compactified Euclidean
space. A useful expression for our purpose is (see, e.g., 
Ref. \cite{Ryb}, pg. 24, Eq. (1.37))
\br
Q &=& -\frac{1}{4\pi^2} \int d^3 {\bf r} \sin^2 \Theta (\nabla \Theta ) \cdot
\epsilon^{abc} n^a (\nabla n^b )\times (\nabla n^c) \no \\
&=& -\frac{1}{2\pi^2} \int d^3 {\bf r} \sin^2 \Theta (\nabla \Theta ) \cdot
\vec H .
\er
Here, if the unit target space three-sphere is spanned by a unit four-vector
$X^\alpha $, $\alpha = 1 ,\ldots ,4$, then the meaning of the target space
coordinates $n^a$, $\Theta$ is
\be
X^a = n^a \sin \Theta \, , \quad a= 1,2,3 \quad , \qquad X^4 =\cos \Theta ,
\ee 
and the compactification condition may be chosen, e.g., 
\be
\lim_{{\bf r}\to \infty} \Theta =\pi ,
\ee
which means that infinity is mapped to the south pole of the three-sphere.
Comparing the two expressions we identify
\be
 g^{(2)} \equiv e^{-\kappa \xi} = \frac{1}{\pi} (\Theta -\frac{1}{2}
\sin 2\Theta ) \quad \Rightarrow \quad
\nabla g^{(2)} =\frac{2}{\pi} \sin^2 \Theta \nabla \Theta
\ee
and the compactification condition becomes
\be
\lim_{{\bf r}\to \infty} g^{(2)} =1
\ee
which is precisely the boundary condition $\xi (r=\infty)=\infty$, see
Eq. (\ref{bound-xi}). 

Therefore, we find
\be
E_{\rm Bog} =\frac{4\pi}{\kappa} Q
\ee
and the Bogomolny energy may indeed be expressed by a topological charge.
In the case of our static solutions of Section 3 it is the field $u$ which
describes a map $S^2 \to S^2$ with winding number $|m|$ and, therefore,
provides the nontrivial winding number $Q=|m|$. With this identification,
expression (\ref{energy}) for the energies of the static configurations  
is exactly equal to the Bogomolny energy, which we wanted to prove. 

\section{The more general system}

We now want to study the case of the more general gauge field (\ref{CFNS-res})
with $C_\mu \not= 0$. The Yang--Mills dilaton lagrangian then reduces to
\be \label{gen-lag-0}
{\cal L} = \frac{1}{4} \left( 2 \partial^\mu \xi \partial_\mu \xi -
e^{-2\kappa \xi}  \hat F^{\mu\nu} \hat F_{\mu\nu} \right) ,
\ee
\be \label{hat-F}
\hat F_{\mu\nu} \equiv F_{\mu\nu} - H_{\mu\nu},
\ee
where
\be
F_{\mu\nu} =\partial_\mu C_\nu - \partial_\nu C_\mu
\ee
is the field strength of the abelian gauge field $C_\mu$ and $H_{\mu\nu}$ is
defined in (\ref{H_munu}). 

\subsection{General remarks}

It may appear from  expression (\ref{gen-lag-0}) 
that the system with $C_\mu \not= 0$
is equivalent to the system with $C_\mu =0$ and may be described by simply
replacing $H_{\mu\nu}$ by $\hat F_{\mu\nu}$, but this is not true. The
important point here is that $H_{\mu\nu}$ as an antisymmetric $4\times 4$
matrix is of second rank. Therefore,
if  $H_{\mu\nu}$ is interpreted as an electromagnetic field tensor, it
corresponds to fields such that the electric and magnetic fields are
perpendicular, $\vec E \cdot \vec B =0$ (the so-called ``radiation fields'').
On the other hand, no condition is imposed on the abelian gauge field $C_\mu$,
therefore $F_{\mu\nu}$ generically is a non-degenerate fourth rank matrix.
Observe that one condition  $\vec E \cdot \vec B =0$ is sufficient to reduce
from a fourth rank matrix to a second rank one. The reason is that the
eigenvalues of an antisymmetric matrix always come in pairs $\pm i\lambda$,
so one rank-reducing condition always sets two eigenvalues equal to zero and
reduces the rank by two. As a consequence, any radiation field may be locally
described by a tensor $H_{\mu\nu}$. Globally, the set of radiation fields
that can be described by a tensor $H_{\mu\nu}$ is more restricted, 
at least as long as
one requires that $n^a$ is a globally well-defined map from $\RR^3$ to $S^2$
(see e.g. \cite{Ran1}). 
 
It is, however, possible to remove the $n$-term in the expression of the gauge
potential by a gauge transformation. 
Indeed, choose the unit vector field $m^a$, then the gauge transformation
\be \label{U-mn}
U = \sqrt{ \frac{1+\cos \gamma }{2}} - \frac{i}{\sqrt{
2(1+\cos\gamma )}} \alpha^a \sigma^a
\ee
\be
\alpha^a \equiv \epsilon^{abc} n^b m^c \, ,\quad \cos\gamma =n^a m^a
\ee
transforms the unit vector field $n^a$ into the new unit vector field $m^a$
(more precisely, the matrix ${\bf n} \equiv n^a \sigma^a$ into the matrix
${\bf m} \equiv m^a \sigma^a$).
If we now choose $m^a =$ const., then the $n$-term is absent in the
gauge-transformed gauge potential, because $m^a_\mu =0$. Specifically, for
$m^a = \hat e_3^a \equiv \delta^{a3} = (0,0,1)$, 
the corresponding gauge transformation 
(\ref{U-mn}) is
\be \label{U-n3}
U= \sqrt{ \frac{1+n^3}{2}} - \frac{i}{\sqrt{2(1+n^3 )}} (n^2 \sigma^1 - n^1
\sigma^2 ).
\ee
This expression is not well-defined at $n^a =(0,0,-1)$ (its value depends on
the order in which the limit $n^1 \to 0$, $n^2 \to 0$ is performed). A
topologically nontrivial $n^a$ covers the whole target space $S^2$ and,
therefore, also the value $(0,0,-1)$. The corresponding $U$ is, therefore,
singular. If the field strength $H_{\mu\nu}$ corresponding to $n^a$ is
regular, however, there always exists a further gauge transformation 
$V=\exp (i\beta \sigma^3 ) $ (for some appropriate $\beta$ not
expressible in terms of $n^a$ only)
which leaves $\hat e_3^a  =(0,0,1)$ invariant, 
such that the composition $VU$ is regular everywhere. (Observe that this does
not happen for the solutions of Section 3, because there the field strength is
the singular magnetic monopole field.)

Next, we want to calculate how a gauge potential transforms under this gauge
transformation. If we define a general Lie-algebra valued gauge potential as
\be
A_\mu = \frac{1}{2}A^a_\mu \sigma^a
\ee
then this gauge potential transforms like
\be
A_\mu \to UA_\mu U^\dagger -i U\partial_\mu U^\dagger
\ee
(the factor $-i$ in front of the inhomogeneous term is due to the fact that
we use the hermitian basis $(1/2) \sigma^a$ in the Lie algebra instead of the
anti-hermitian $t^a =-(i/2) \sigma^a$). 
Specifically, if we choose for the untransformed gauge potential
\be 
A^a_\mu = \epsilon^{abc} n^b_\mu n^c
\ee
consisting only of the $n$-term, and the transformation (\ref{U-n3}) for the
gauge transformation, then the transformed gauge potential is
\be
A'^a_\mu  = \frac{n^2_\mu n^1 - n^1_\mu n^2}{1+n^3} \hat e^a_3
\ee
that is, only the $C_\mu$-term is nonzero, and the transformed unit vector is
$m^a  = \hat e_3^a$. If this gauge potential is singular (because $n^a$ is
topologically nontrivial) but leads to a non-singular fields strength, then
there always exists a further gauge transformation (the above-mentioned $V$) 
which transforms  $ A'^a_\mu$ to a regular gauge potential. This
transformation acts as an abelian gauge transformation on the abelian gauge
field $C_\mu$, i.e.
\be
V \, : \quad \frac{n^2_\mu n^1 - n^1_\mu n^2}{1+n^3} \quad \to \quad
\frac{n^2_\mu n^1 - n^1_\mu n^2}{1+n^3} + \beta_\mu .
\ee

Now we want to discuss the conservation laws of the Lagrangian
(\ref{gen-lag-0}). For this purpose it is useful to rewrite it like
 (remember $c\equiv (u^\mu \bar u_\mu )^2 - 
(u_\mu)^2 (\bar u_\nu)^2 $)
\be \label{gen-lag-1}
{\cal L} = \frac{1}{2}   \partial^\mu \xi \partial_\mu \xi - 
e^{-2\kappa \xi} \left( 
 \frac{2c}{
(1+u\bar u)^4} - 
i \frac{u_\mu \bar u_\nu - u_\nu \bar u_\mu}{(1+u\bar u)^2}F^{\mu\nu} +
\frac{1}{4}F^{\mu\nu}F_{\mu\nu} \right) .
\ee
This Lagrangian still has the infinitely many conserved currents
(\ref{J-cal-G}), because the term $(1+u\bar u)^{-2} (u_\mu \bar u_\nu - u_\nu
\bar u_\mu )$ is still invariant under the corresponding target space
transformations $\delta u = i (1+u\bar u)^2 {\cal G}_{\bar u}$, etc., as may
be checked easily. But the above Lagrangian has even more conserved currents,
as we demonstrate now. For this purpose, we rewrite it in the slightly more
general form  
\be \label{gen-lag-2}
{\cal L} = \frac{1}{2}   \partial^\mu \xi \partial_\mu \xi - 
G^{(2)}(\xi) \left( 
2 G^{(1)}(a) c - 
i g^{(1)} (a) (u_\mu \bar u_\nu - u_\nu \bar u_\mu )F^{\mu\nu} +
\frac{1}{4}F^{\mu\nu}F_{\mu\nu} \right) 
\ee
like in (\ref{en-int-mod-gen}), (remember $G^{(1)} \equiv (g^{(1)})^2$), and
calculate the following Euler--Lagrange equations, for $\bar u$
\be \label{gen-eq-u}
2\partial_\mu \left( G^{(2)} g^{(1)} K^\mu \right) +i g^{(1)} \partial_\mu
\left( G^{(2)} F^{\mu\nu} u_\nu \right) =0 ,
\ee
for $\xi$,
\be \label{gen-eq-xi}
\xi^\mu_\mu + G^{(2)}_\xi \left( 2 G^{(1)}(a) c - 
i g^{(1)} (a) (u_\mu \bar u_\nu - u_\nu \bar u_\mu )F^{\mu\nu} +
\frac{1}{4}F^{\mu\nu}F_{\mu\nu} \right) =0 ,
\ee
for $C_\mu$, 
\be \label{gen-eq-C}
\partial^\mu \left[ G^{(2)} \left( F_{\mu\nu} -2i g^{(1)} 
(u_\mu \bar u_\nu - u_\nu \bar u_\mu ) \right) \right] =0. 
\ee
It follows easily from the last field equation that the currents    
\be
 j_{\mu}^{\cal H}= {\cal H}(\xi) G^{(2)} (\xi) \left( F_{\mu\nu} -2i g^{(1)} 
(u_\mu \bar u_\nu - u_\nu \bar u_\mu ) \right) \xi^{\nu}
\ee
are conserved, where ${\cal H}(\xi)$ is an arbitrary real 
function of its argument. These currents are completely analogous to the 
additional conserved currents which were found for abelian gauge theories in
\cite{int-U1}, see Eqs. (61) - (63) of that paper.   

Remark: Setting $u=0$ and observing that $F_{\mu\nu}$ is a general
abelian gauge field strength, it follows that the general Maxwell dilaton
system has infinitely many conserved currents.

\subsection{Further solutions}
A first possibility to construct further solutions is provided by just adding
a pure gauge $C$-term, i.e. a term $(\partial_\mu \lambda ) n^a$ to a
given solution. This is a trivial modification from the point of view of the
integrable submodel, because its Lagrangian only depends on the field strength
$F_{\mu\nu}$ (see (\ref{gen-lag-0}), but it is a nontrivial modification from
the point of view of the full gauge potential (with non-zero $W$-term),
because the Lagrangian of the full theory depends on the abelian gauge
potential $C_\mu$ (and not only on the abelian field strength). 
The term  $(\partial_\mu \lambda ) n^a$ can be removed by a
nonabelian gauge transformation (with gauge parameter $\alpha^a = \lambda
n^a$) for a general gauge potential, but this transformation acts nontrivially
on the $W$-term and so just transfers the physical effect of  
$(\partial_\mu \lambda ) n^a$ from the $C$-term to the $W$-term.
 
Specifically, by choosing $\lambda =q x^4$ (here $q$ is a constant, $x^4$ is
Euclidean ``time'', and we momentarily switch to Euclidean space-time
conventions to be consistent with Ref. \cite{BriLa1}), 
i.e., by adding a $C$-term with
$C_\mu = q \delta_{\mu 4}$ to the monopole type solutions of Section 3, we are
able to reproduce the special analytic solution of Ref. \cite{BriLa1}
(they are called ``dyonic-type generalisations of the monopole solutions'' in
that paper, see their Eq. (17); for the constant $q$ they use the symbol $\bar
u$, which we do not employ here for obvious reasons). Obviously, these
solutions have the same energies and fulfill the same Bogomolny bounds as the
solutions of Section 3.

We may try to find further solutions by solving the full system of field 
equations (\ref{gen-eq-u}) - (\ref{gen-eq-C}). We assume that no field
variable depends on time (static solutions), and we further assume that
$C_\mu$ only describes an electric field (i.e., $F_{0j}=E_j$, 
$F_{jk}=0$; observe that 
$H_{\mu\nu} \to H_{jk}$ is automatically purely magnetic for static $u$). 
Under these assumptions, the second term in the field equation for $\bar u$,
Eq. (\ref{gen-eq-u}), does not contribute, and we get Eq. (\ref{EL-u}) of
Section 3. Also the term $(u_\mu \bar u_\nu - u_\nu \bar u_\mu ) F^{\mu\nu}$
in Eq. (\ref{gen-eq-xi}) is zero under these assumptions. Now we further
restrict to
\be \label{ansatz-2}
u=v(\theta) e^{im\varphi} \, , \quad \xi = \xi(r) \, , \quad 
\vec E = E(r) \hat e_r
\ee
and find for $u$ exactly the same solutions as in Section 3 (i.e., the first
integral (\ref{first-int-theta}) for general $g^{(1)}$, and the generalised
hedge-hogs (\ref{sol-theta}) for $g^{(1)} =(1+u\bar u)^{-2}$). For these
solutions for $u$, the second term in Eq. (\ref{gen-eq-C}) is zero, and 
Eq. (\ref{gen-eq-C})   simplifies to
\be
\nabla (G^{(2)} E(r) \hat e_r )=0
\ee
with the solution
\be \label{el-sol-C}
E(r)= \frac{q}{4\pi r^2 G^{(2)}}
\ee
where $q$ is the constant electrical charge.  Finally, Eq. (\ref{gen-eq-xi})
leads to the first integral
\be
\xi_s^2 =16 m^2 \mu^2 G^{(2)} + \frac{q^2}{4\pi^2 G^{(2)}} +\tilde \lambda
\ee 
where $s\equiv (1/r)$ and the calculation is completely analogous to the
calculation in Section 3. For a further evaluation we have to specify
$G^{(2)}(\xi)$. For the dilaton case $G^{(2)} = \exp (-2\kappa \xi )$ and the
generalised hedge-hogs for $u$ we get (remember $\mu^{-1}=4$)
\be
\xi_s^2 = m^2  e^{-2\kappa \xi } + \frac{q^2}{4\pi^2 } e^{2 \kappa \xi}
+\tilde \lambda .
\ee
Unfortunately, there does not exist a physically acceptable (finite energy)
solution if both $m$ and $q$ are different from zero. Choosing $q=0$ we
recover the solutions of Section 3, but choosing $m=0$, we find different,
purely electric solutions. Finite energy requires again $\tilde \lambda =0$,
and we get
\be
\xi_s^2 =  \frac{q^2}{4\pi^2 } e^{2 \kappa \xi} .
\ee
Comparing with the equation
\be
\xi_s^2 =  m^2 e^{-2 \kappa \xi} 
\ee
for the magnetic solutions of Section 3, we see that we recover the electric
equation by the replacements $m \to (q/2\pi )$ and $\xi \to -\xi$, therefore
the electric solution for $\xi$ is
\be \label{el-sol-xi}
\xi = - \frac{1}{\kappa } \ln \left( \frac{ |q|}{2\pi }\kappa s +1 \right) .
\ee
Also the energy of the electric solution (\ref{el-sol-C}),
(\ref{el-sol-xi}) is the same as the energy of the magnetic solution
(\ref{energy}) after the replacement $m \to  (q/2\pi )$. The electric
solutions are also Bogomolny solutions, and the Bogomolny bound may be derived
exactly like in Section 4, where one only has to replace the ``magnetic''
vector $\vec H$ by the electric vector $\vec E =- \nabla C_0$.  

Remark: there does not exist a Bogomolny bound when both electric and magnetic
vectors are present.

Remark: the electric solution (\ref{el-sol-C}), (\ref{el-sol-xi}) has been
derived in \cite{Biz} from the simplest magnetic solution in a slightly 
different way, by employing the nontrivial duality symmetry
\be
F^a_{\mu\nu} \to e^{-2\kappa \xi} \tilde F^a_{\mu\nu} \, ,\quad
\xi \to -\xi
\ee
which is present for the Yang--Mills dilaton theory 
(here $\tilde F^a_{\mu\nu} $ is the
Hodge dual of the nonabelian field strength $F^a_{\mu\nu}$).  

\subsection{Further generalisations}
Finally, let us briefly discuss some generalisations of the integrable
Lagrangians discussed so far, which are still integrable (have infinitely many
symmetries) and allow for the ansatz (\ref{ansatz}) in spherical polar
coordinates, and for trivial first integrals of the resulting  ODEs. One class
of models is given by the Lagrangians (with non-polynomial kinetic terms)
\be
{\cal L}= \frac{1}{2} (\xi_{\mu}\xi^{\mu})^\alpha - 2
G^{(2)}(\xi)G^{(1)}(a) ((u^\mu \bar u_\mu )^2 - 
(u_\mu )^2 ( \bar u_\nu )^2 )^\beta
\ee
where $\alpha ,\beta$ are parameters. This model has the infinitely many
conserved currents (analogously to the currents (\ref{J-cal-G}))
\be \label{J-cal-G-gen}
J^{ \cal G}_\mu = i (G^{(1)})^{-\frac{\beta}{2}} 
[{\cal G}_{\bar u} \pi_\mu - {\cal G}_u 
\bar \pi_\mu ]
\ee
where ${\cal G}={\cal G}(u,\bar u)$ 
is an arbitrary real function of its arguments. Further, the separation of
variables ansatz (\ref{ansatz}) is compatible with the Euler--Lagrange
equations, and the resulting ODEs are solvable by quadratures. Whether there
exist finite energy static solutions depends both on the choice of the
parameters $\alpha ,\beta$ (where Derrick's theorem provides a selection
criterion) and on the functions $G^{(1)}$, $G^{(2)}$.
In the special case $\alpha =\frac{3}{2}$, $\beta = \frac{3}{4}$, the energy
functional for static solutions enjoys an enhanced base space symmetry
in $\RR^3$ (conformal symmetry instead of just Galilean symmetry) and,
therefore, a separation of variables ansatz in toroidal coordinates
is compatible with the field
equations and leads to static solutions, quite analogously to the case of the 
model of Aratyn, Ferreira, and Zimerman \cite{AFZ1}, \cite{AFZ2}, \cite{BF1}.

Another generalisation is given by
\br 
{\cal L} &=& \frac{1}{2}   \partial^\mu \xi \partial_\mu \xi - 
2 G^{(2)}(\xi) 
2 G^{(1)}(a) c - \nonumber \\ && -
i H^{(2)}(\xi ) H^{(1)} (a) (u_\mu \bar u_\nu - u_\nu \bar u_\mu )F^{\mu\nu} +
\frac{1}{4}K^{(2)}(\xi )  F^{\mu\nu}F_{\mu\nu}  .\label{gen-lag-3}
\er
If $G^{(1)} = (H^{(1)})^2$, then this Lagrangian has the infinitely many
conserved currents (\ref{J-cal-G})); if this condition does not hold, only the 
smaller set of currents (\ref{J-tilde-G}) is conserved. Further, the currents
\be
j_{\mu}^{\cal H}= {\cal H}(\xi) \left( K^{(2)} (\xi)  F_{\mu\nu} -2i H^{(1)} 
H^{(2)} (u_\mu \bar u_\nu - u_\nu \bar u_\mu ) \right) \xi^{\nu}
\ee
are conserved, where ${\cal H}(\xi)$ is an arbitrary real 
function of its argument. The separation of
variables ansatz (\ref{ansatz-2}) for static solutions and a purely electric
$F_{\mu\nu}$  is again compatible with the Euler--Lagrange
equations, and the resulting ODEs are again solvable by quadratures.
The existence of finite energy static solutions depends again on the choice of
the arbitrary functions. For instance,
there exist infinitely many solutions with both  non-zero magnetic and
electric fields for the following choice,
\be
G^{(1)} =(1+a)^{-4} \, ,\quad G^{(2)} =(K^{(2)})^{-1}
=e^{-2\kappa \xi } .
\ee
Here the only difference to the Yang--Mills dilaton case is that 
$(K^{(2)})^{-1}=e^{-2\kappa \xi }$ instead of $K^{(2)} =e^{-2\kappa \xi }$
(the values for $H^{(1)}$, $H^{(2)}$ are irrelevant because the term 
multiplying them does
not contribute for purely magnetic $H_{\mu\nu}$ and purely electric 
$F_{\mu\nu}$).

\section{Discussion}

It has been one of the initial motivations 
of this investigation to shed more light on
previous results on the Yang--Mills dilaton theory. In this respect, our first
result is the simple observation that Yang--Mills dilaton theory contains an
integrable subsystem, i.e., a subsystem with infinitely many target space
symmetries and infinitely many conserved currents, and 
this subsystem is non-empty in the sense that it contains nontrivial (e.g.,
static finite energy) solutions. This fact is also
intimately related to our second result, namely a possible explanation for the
existence of infinitely many static analytic solutions. In this context, it
is interesting to note the role which is played by the symmetries of the
integrable subsystem. The consistency of the ansatz (\ref{ansatz}) 
in spherical polar coordinates is 
explained by the base space symmetries of the model (essentially by rotational
symmetry). On the other hand, the solvability of the resulting ODEs by simple
quadratures might be related to the
 integrability of the model, i.e., to the existence of infinitely many
target space symmetries and conservation laws. The conjecture
that solvability follows from integrability is supported both by the
corresponding, rigorous results in lower dimensions and by the fact that it is
true in all known examples of higher-dimensional integrable theories.
It holds, e.g., for the model of Aratyn, Ferreira, and Zimerman
\cite{AFZ1}, \cite{AFZ2}
where the base space symmetries allow for an ansatz in toroidal coordinates
such that the resulting ODEs are solvable by quadratures, 
or for a class of models similar to the one of this paper, but
with non-polynomial kinetic energy expressions, see \cite{ASGW5}. A more
rigorous mathematical analysis of the relation between integrability and 
solvability for integrable theories in higher dimensions is still missing 
and would be highly desireable. 

At this point a word of caution may be appropriate. Although the relation
between infinitely many conservation laws and (at least classical) solvability
may carry over from 1+1 to higher dimensions, this is certainly not the case
for some other features of 1+1 dimensional integrable theories. In 1+1
dimensional integrable theories one has, for instance, the possibility of
conserved charges of arbitrary spin which, in turn, lead to the factorizability
of the S-matrix. This cannot happen in higher dimensions due to the 
Coleman--Mandula theorem \cite{CoMa}. 
Also the transition from classical to quantum
integrable systems will most likely be more complicated in higher dimensions.
The algebra of the infinitely many conserved charges or conserved currents,
as exposed e.g. in \cite{vol-pres}, \cite{ASG4}, \cite{ASGW1} for certain
classes of theories, may, nevertheless, be a good starting point for the
quantization. This issue is, however, beyond the scope of the present article.

Our third result is the explanation of the fact that 
the energies of our static solutions
grow linearly with the topological charge. This is explained by the Bogomolny
bound (\ref{bogol-ineq}) for the integrable submodel and by the observation
that the static solutions saturate this bound.

Remark: for the simplest solution with $m=1$, a Bogomolny type bound
was given in \cite{Biz}. However, this Bogomolny bound holds for a 
certain subsector
of the spherically symmetric ansatz of the Yang--Mills dilaton system, 
therefore it only applies to spherically symmetric solutions. On the
other hand, our Bogomolny bound (\ref{bogol-ineq}) holds completely
generally for the integrable subsystem and does not require a certain symmetry
of the solution. 

In Section 5 we found similar results for a slightly more general Lagrangian
(i.e. for a slightly less restricted gauge potential) still maintaining
integrability and solvability. It is interesting that both finite energy 
solutions and Bogomolny bounds seem to exist only for purely magnetic or
purely electric gauge fields, but not for the mixed case. Further, we
presented some more general Lagrangians, no longer related to the Yang--Mills
dilaton system, which are still both integrable and solvable.

Finally let us point out that,
apart from sheding more light on some issues of 
Yang--Mills dilaton theory, our investigation has an additional interest. 
Observe that we have used the decomposition of Cho, Faddeev, Niemi, and
Shabanov - which originally was mainly motivated by the description of the
low-energy degrees of freedom of Yang--Mills theory - for a different
purpose, namely for the exposure of an intergrable subsector within this
theory. Here it is important to note that the target space transformations -
which are symmetry transformations in the submodel - are very simple
in terms of the decomposition fields. They are essentially
area-preserving diffeomorphisms of the target space two-sphere spanned by the 
unit vector $n^a$ (or by the complex field $u$), therefore they are just
geometric transformations. On the other hand, in terms of the
original Yang--Mills field, they certainly would be rather nontrivial,
nonlocal transformations, which would be quite difficult to detect.  
  
This immediately rises the question of generalisations and further
applications. First of all, we believe that most likely our investigation
carries over without problems to the case of the Einstein--Yang--Mills dilaton
system (that is, to an appropriately chosen subsystem thereof), given the
striking similarity between the two theories found, e.g. in \cite{Biz}. 
Another interesting question is whether the CFNS decompostion or a
similar decomposition can be used, e.g., to unravel an integrable subsector to
which the sector of self-dual solutions of Yang--Mills theory
belongs. These problems are under
investigation. In any case, it seems that the use of nonlocal field
transformations is a useful instrument for the discovery of nontrivial
symmetries and integrable subsectors of nonlinear field theories.  

\section*{Appendix}
Here we want to prove that inserting the CFNS decomposition (\ref{CFNS-dec})
directly into the Yang--Mills dilaton
action and deriving the Euler--Lagrange equations with
respect to the decomposition fields gives the same field equations as
the ones that are
obtained by inserting the decomposition into the Euler--Lagrange equations of
the original Yang--Mills dilaton theory. 
For the case of pure SU(2) Yang--Mills theory this equivalence of the
two different ways to derive the field equations was 
already proven in \cite{FN1}, but as we study a different theory 
in our paper and use a slightly different parametrization for the
decomposition, we provide the proof here for the convenience of the reader.

The dilaton field is not changed
under the decomposition, therefore the equivalence obviously holds for the 
dilaton
field equation. So we focus on the Yang--Mills equation in the sequel.
In terms of the Yang--Mills gauge potential $A^a_\mu$, the corresponding 
Euler--Lagrange equation reads
\be \label{YM-eq}
e^{-2\kappa \xi } \epsilon^{abc} A^b_\nu F^{c\nu\lambda} +
\partial_\nu (e^{-2\kappa \xi} F^{a\nu \lambda })=0.
\ee  
In a first step we want to evaluate this expression for the gauge invariant
abelian projection, that is, the restriction (\ref{CFNS-res}) with both 
$n^a$ and $C_\mu$ nonzero. Inserting this decomposition we easily find the
equation
\be \label{YM-Cu}
n^a \partial_\mu (e^{-2\kappa \xi} \hat F^{\mu\nu})=0
\ee
(where $\hat F^{\mu\nu}$ is defined in (\ref{hat-F})), 
and the factor of $n^a$ in
front of the expression is obviously irrelevant and may be omitted.
This now has to be compared with Eqs. (\ref{gen-eq-u}) and (\ref{gen-eq-C})
for the special case $G^{(2)}=e^{-2\kappa \xi}$, $g^{(1)} = (1+u\bar u)^{-2}$.
Equation (\ref{gen-eq-C}) may be written like
\be \label{gen-eq-C-1}
\partial_\mu (e^{-2\kappa \xi} \hat F^{\mu\nu})=0
\ee
and is, therefore, identical to the above Equation (\ref{YM-Cu}). 
Equation (\ref{gen-eq-u}) may be written like
\be \label{gen-eq-u-1}
u_\nu \partial_\mu (e^{-2\kappa \xi} \hat F^{\mu\nu})=0
\ee
together with its complex conjugate, and is, therefore, just the projection 
of Eq. (\ref{YM-Cu}) into the directions of $u_\mu$ and $\bar u_\mu$.
Obviously, Eqs. (\ref{YM-Cu})  are completely equivalent to
Eqs. (\ref{gen-eq-C-1}), (\ref{gen-eq-u-1}). 

Finally, let us discuss what happens when we further restrict to
(\ref{CW-restr}), that is, we set $C_\mu =0$, too. This produces a small
complication that is related to the fact that this latter restriction is no
longer gauge invariant. Inserting this restriction into the
Yang--Mills equation (\ref{YM-eq}) simply gives
\be \label{YM-u}
n^a \partial_\mu (e^{-2\kappa \xi}  H^{\mu\nu})=0
\ee  
whereas variation of the Lagrangian (\ref{int-lag}) w.r.t. $\bar u$ gives,
in complete analogy with (\ref{gen-eq-u-1})
\be \label{gen-eq-u-2}
u_\nu \partial_\mu (e^{-2\kappa \xi} H^{\mu\nu})=0.
\ee
Together with its complex conjugate, this seems to provide only two equations
(the projections into the directions of $u_\mu$ and $\bar u_\mu$) 
compared to the four equations (\ref{YM-u}). This apparent mismatch is,
however, easily understood and is related to the fact that the restriction
to $C_\mu =0$ is no longer gauge invariant. The original Yang--Mills equation
is gauge covariant and leads, after inserting the restricted decomposition,
to the gauge covariant field equations (\ref{YM-u}). Because of the abelian
character of the gauge field, these equations are in fact gauge invariant 
after omission of the irrelevant factor $n^a$. On the other hand, setting
$C_\mu =0$ already in the Lagrangian involves a gauge choice, and the
resulting field equations (\ref{gen-eq-u-2}) hold only in this gauge.
A gauge variation of these field equations automatically switches on the two
missing components.  This may be seen especially easily when the Lagrangian
of the submodel is varied w.r.t. $n^a$ instead of $u$, because $n^a$ has
a simple behaviour under gauge transformations. Variation of the Lagrangian
(\ref{int-lag-0}) w.r.t. $n^a$ leads to
\be \label{gen-eq-n-1}
\epsilon^{abc} n^b_\mu n^c \partial_\lambda (e^{-2\kappa \xi } H^{\mu\lambda }
)=0
\ee
which is identical to Eq. (\ref{gen-eq-u-2}) when expressed in terms of $n^a$
instead of $u$. Again, Eq. (\ref{gen-eq-n-1}) only consists of two components.
Further, Eq.  (\ref{gen-eq-n-1}) consists of the gauge invariant factor
\be
\partial_\lambda (e^{-2\kappa \xi } H^{\mu\lambda })
\ee
and the gauge dependent prefactor
\be
\epsilon^{abc} n^b_\mu n^c .
\ee
Under a gauge variation $\delta n^a = \epsilon^{abc}n^b \alpha^c$ this factor
changes according to
\be
\delta ( \epsilon^{abc} n^b_\mu n^c ) = (\delta^{ab} - n^a n^b )\alpha^b_\mu
+ (n^a n^b_\mu - n^a_\mu n^b )\alpha^b .
\ee
As the gauge variation $\alpha^a (x)$ is completely arbitrary, already the
first term proportional to $\alpha^a_\mu$ contains, in general, projections
onto the two missing components of the Yang--Mills field equations.
\\ \\ \\
{\large \bf Acknowledgement} \\
 C.A. and J.S.-G. thank MCyT (Spain) and FEDER
(FPA2005-01963), and support from 
 Xunta de Galicia (grant PGIDIT06PXIB296182PR and Conselleria de
Educacion).
A.W. is
partially supported from Adam Krzy\.{z}anowski Fund. Further, C.A.
acknowledges support from the Austrian START award project
FWF-Y-137-TEC and from the FWF project P161 05 NO 5 of N.J.
Mauser. Finally, A.W. thanks P. Bizon for helpful discussions.

\end{document}